\documentclass[structabstract]{aa}
\usepackage{bbding}
\usepackage{pifont}
\usepackage{amsmath}
\usepackage{graphicx}
\usepackage{stmaryrd}
\usepackage{txfonts}

\usepackage{url}
\usepackage{subfigure}
\usepackage{longtable}
\usepackage{lscape}
\usepackage{natbib}
\begin{document}
   \title{Molecular clumps and star formation associated with the infrared dust bubble N131}

    \authorrunning{C. P. Zhang et al.}
    \titlerunning{The bubble N131}

   \author{Chuan-Peng Zhang
          \inst{1,2,3}
          \and
          Jun-Jie Wang\inst{1,2}
          \and
          Jin-Long Xu\inst{1,2}
          }

   \institute{National Astronomical Observatories, Chinese Academy of Sciences,
              100012 Beijing, PR China\\
             \email{zcp0507@gmail.com}
         \and
             NAOC-TU Joint Center for Astrophysics, 850000 Lhasa, PR China
         \and
             University of the Chinese Academy of Sciences, 100080 Beijing, PR China
             }

   \date{Received ---; accepted ---}


  \abstract
   {}
{The aim is to explore the interstellar medium around the dust
bubble N131 and search for signatures of star formation.}
{We have performed a multiwavelength study around the N131 with data
taken from large-scale surveys of infrared observation with online
archive. We present new observations of three CO $J = 1 - 0$ isotope
variants from Purple Mountain Observatory 13.7 m telescope. We
analyzed the distribution of the molecular gas and dust in the
environment of the N131. We used color-color diagrams to search for
young stellar objects and to identify ionizing star candidates.}
{The kinematic distance of $\sim$8.6 kpc has been adopted as the
distance of the bubble N131 from the Sun in this work. We find a
ring of clouds in CO emission coincident with the shell of N131 seen
in the Spitzer telescope images, and two giant elongated molecular
clouds of CO emission appearing on opposite sides of the ringlike
shell of N131. There is a cavity within the bubble at 1.4 GHz and 24
$\mu$m. Seven IRAS point sources are distributed along the ringlike
shell of the bubble N131. Fifteen ionizing stars and 63 YSO
candidates have been found. The clustered class I and II YSOs are
distributed along the elongated clouds in the line of sight.}
    {}


   \maketitle
%

\section{Introduction}    
\label{sect:intro}


\citet{chur2006,chur2007} have detected and cataloged about 600
midinfrared dust (MIR) bubbles between longitudes -60$^{\circ}$ and
+60$^{\circ}$. The IR dust bubbles may be produced by ionizing O-
and/or B-type stars, which are located inside the bubble. The
ultraviolet (UV) radiation from ionizing stars may heat dust and
ionize the gas to form an expanding bubble shell \citep{wats2008}.
\citet{simp2012} present a new catalog of 5106 infrared bubbles
created through visual classification via the online citizen science
website ``The Milky Way
Project"\footnote{http://www.milkywayproject.org}, which provides a
crowd-sourced map of bubbles and arcs in the Milky Way, and will
enable better statistical analysis of Galactic star-forming sites.
\citet{beau2010} report CO J = 3 - 2 maps of 43 Spitzer identified
bubbles, which suggests that expanding shock fronts are poorly bound
by molecular gas and that cloud compression by these shocks may be
limited. \citet{wats2008} present an analysis of wind-blown,
parsec-sized, midinfrared bubbles, and associated star formation,
and suggest that more than a quarter of the bubbles may have
triggered the formation of massive objects.

A few individual bubbles have been studied well, such as N22
\citep{jiwg2012}, N49 \citep{wats2008,deha2010,zava2010}, N65
\citep{petr2010}, N68 \citep{n68}, and S51 \citep{s51}. There are
many models and observations to explain the dusty wind-blown
bubbles, such as bubble N49 of \citet{ever2010}. Recently, we have
reported an expanding ringlike shell of the bubble S51, which shows
a rare front side located within the shell in the line of sight, by
employing $^{13}$CO and C$^{18}$O J = 1 - 0 emission lines of the
Mopra Telescope \citep{s51}. We also investigated the star formation
around the bubble N68, which suggests that the massive star
formation at the ringlike shell is very active \citep{n68}.

\begin{table*}
\caption{Parameters of the nine spectra} \label{table:spectra}
\centering \scriptsize
\begin{tabular}{ccccccccccccc}
\hline \hline
Clump  &    R.A.(J2000)  &  DEC.(J2000)  &  V$\rm{_{^{12}CO}}$ &  V$\rm{_{^{13}CO}}$  & $\Delta$V$\rm{_{^{12}CO}}$ &  $\Delta$V$\rm{_{^{13}CO}}$  & T$\rm{_{^{12}CO}}$ &   T$\rm{_{^{13}CO}}$   \\
      &  $^{h}$~~~$^{m}$~~~$^{s}$ & $^{\circ}$~~~$'$~~~$''$   & (km s$^{-1}$) & (km s$^{-2}$) & (km s$^{-1}$) & (km s$^{-2}$)   & (K) & (K)    \\
(1)   &  (2)   &  (3)   &  (4)   &  (5)   &  (6)   &  (7)   &  (8) &  (9)  \\

\hline
A     &    19 51 55.681   &   26 21 29.87  &      -8.91$\pm$0.01   &   -9.25$\pm$0.03   &  2.63$\pm$0.02  &  2.14$\pm$0.06  & 11.40$\pm$0.25  &   2.04$\pm$0.13    \\
B     &    19 52 45.568   &   26 25 54.58  &     -11.89$\pm$0.02   &  -12.15$\pm$0.03   &  2.65$\pm$0.05  &  2.05$\pm$0.07  &  7.19$\pm$0.31  &   2.38$\pm$0.16    \\
C     &    19 53 01.891   &   26 28 06.93  &     -12.24$\pm$0.04   &  -12.53$\pm$0.08   &  2.62$\pm$0.09  &  2.53$\pm$0.17  &  4.13$\pm$0.26  &   0.95$\pm$0.16    \\
D     &    19 51 32.383   &   26 20 22.06  &      -9.93$\pm$0.02   &  -10.15$\pm$0.10   &  2.30$\pm$0.05  &  2.23$\pm$0.29  &  6.07$\pm$0.34  &   0.95$\pm$0.19    \\
E     &    19 52 18.309   &   26 27 02.22  &      -8.09$\pm$0.03   &   -8.29$\pm$0.05   &  1.55$\pm$0.09  &  1.63$\pm$0.17  &  5.22$\pm$0.23  &   1.02$\pm$0.16    \\
F     &    19 52 51.126   &   26 19 23.40  &      -9.80$\pm$0.03   &   -9.84$\pm$0.12   &  1.77$\pm$0.06  &  2.05$\pm$0.37  &  5.25$\pm$0.36  &   0.62$\pm$0.16    \\
G     &    19 52 35.316   &   26 18 01.05  &     -10.64$\pm$0.01   &  -10.68$\pm$0.02   &  1.97$\pm$0.03  &  1.33$\pm$0.05  &  7.35$\pm$0.31  &   2.24$\pm$0.12    \\
H     &    19 52 16.594   &   26 15 45.75  &     -12.15$\pm$0.03   &  -12.36$\pm$0.10   &  1.82$\pm$0.08  &  1.21$\pm$0.22  &  2.96$\pm$0.27  &   0.50$\pm$0.11    \\
I     &    19 52 02.983   &   26 16 21.05  &      -8.31$\pm$0.04   &   -8.33$\pm$0.09   &  3.85$\pm$0.10  &  2.72$\pm$0.28  &  3.15$\pm$0.29  &   0.73$\pm$0.15    \\
\hline
\end{tabular}
\end{table*}

\begin{table*}
\caption{Derived parameters of the CO molecular clumps}
\label{table:clump} \centering \scriptsize
\begin{tabular}{cccccccccc}
\hline \hline
Clump  &    Area    &   Int.T$\rm{_{^{12}CO}}$ &    Int.T$\rm{_{^{13}CO}}$ &    T$\rm{_{ex}(max)}$  &   T$\rm{_{ex}(mean)}$   &   N$\rm{_{H_{2}}(max)}$   &   N$\rm{_{H_{2}}(mean)}$  &  M$\rm{_{H_{2}}}$ & n$\rm{_{H_{2}}}$  \\
      &   (arcmin$^2$)   & (arcmin$^{2}$$\cdot$K) & (arcmin$^{2}$$\cdot$K)    & (K) & (K)   &  (10$^{21}$ cm$^{-2}$) &  (10$^{21}$ cm$^{-2}$)  & (M$_{\odot}$) & (cm$^{-3}$) \\
(1)   &  (2)   &  (3)   &  (4)   &  (5)   &  (6)   &  (7)   &  (8) &  (9)  &¡¡(10) \\

\hline

AD    &    36.59   &    554.34    &  102.31  &   29.15   &   11.54  &  15.39  &  3.36   &  1438  &  72    \\
BC    &    32.23   &    491.05    &  122.32  &   23.17   &   12.42  &  21.10  &  4.71   &  1775  &  107   \\
A     &    13.72   &    305.22    &  50.05   &   29.15   &   14.56  &  15.39  &  5.07   &  814   &  177   \\
B     &    12.63   &    233.59    &  58.97   &   23.17   &   14.73  &  21.10  &  5.69   &  841   &  207   \\
C     &    6.75    &    118.92    &  30.93   &   21.94   &   17.12  &  17.12  &  8.48   &  669   &  423   \\
D     &    2.18    &    37.48     &  6.66    &   20.90   &   15.12  &  8.90   &  5.00   &  127   &  439   \\
E     &    3.70    &    31.23     &  7.71    &   17.10   &   9.28   &  3.71   &  1.25   &  54    &  84    \\
F     &    3.05    &    40.64     &  7.53    &   17.79   &   12.90  &  4.88   &  2.12   &  76    &  157   \\
G     &    3.70    &    50.50     &  13.38   &   18.95   &   11.19  &  8.42   &  2.53   &  109   &  170   \\
H     &    1.31    &    10.01     &  2.41    &   11.89   &   8.28   &  1.75   &  0.58   &  9     &  66    \\
I     &    2.18    &    27.49     &  5.15    &   13.08   &   8.03   &  3.35   &  1.06   &  27    &  93    \\

\hline
\end{tabular}
\end{table*}

\begin{table*}
\caption{IRAS point sources around the bubble N131}
\label{table:IRAS} \centering \scriptsize
\begin{tabular}{cccccccccc}
\hline \hline
Name & IRAS Source   &   R.A.(J2000)  &   DEC.(J2000)   & $F_{12}$ & $F_{25}$ &  $F_{60}$ & $F_{100}$ & $L_{IR}$ & $T_d$      \\
& & $^{h}$~~~$^{m}$~~~$^{s}$ & $^{\circ}$~~~$'$~~~$''$ & Jy & Jy & Jy & Jy &  $L_{\odot}$ & K                            \\
(1)         &   (2)          &   (3)           & (4)      & (5)      &  (6)      & (7)       &  (8)   &   (9)  &  (10)   \\
\hline
IR1  &  19498+2621  &   19 51 58.18  &   26 29 46.36   &      0.31  &  0.43   &    3.14   &  33.82   &  1855   & 19.47   \\
IR2  &  19499+2613  &   19 52 01.47  &   26 21 09.56   &      5.54  &  22.02  &   216.90  &  322.80  &  40873  & 32.52   \\
IR3  &  19500+2617  &   19 52 09.90  &   26 24 56.10   &      0.45  &  0.27   &    3.23   &  322.80  &  13199  & 13.41   \\
IR4  &  19501+2607  &   19 52 12.62  &   26 15 03.28   &      0.44  &  1.12   &    9.84   &  31.69   &  2627   & 25.78   \\
IR5  &  19502+2606  &   19 52 19.94  &   26 14 19.75   &      4.14  &  1.70   &    2.99   &  31.51   &  3717   & 19.56   \\
IR6  &  19502+2618  &   19 52 21.28  &   26 26 41.84   &      0.28  &  0.27   &    2.55   &  53.40   &  2520   & 17.16   \\
IR7  &  19503+2617  &   19 52 26.01  &   26 25 28.14   &      5.96  &  2.60   &    2.72   &  43.68   &  5149   & 18.01   \\
IR8  &  19505+2610  &   19 52 40.28  &   26 18 04.06   &      0.30  &  0.22   &    3.38   &  23.33   &  1429   & 21.40   \\

\hline
\end{tabular}

\end{table*}

\begin{table*}
\caption{Ionizing star candidates within bubble}
\label{table:ionizing} \centering \scriptsize
\begin{tabular}{ccccccccccccc}
\hline \hline

GLIMPSE Desig.     &  $m_{3.6}$  &  $m_{4.5}$  &  $m_{5.8}$ &  $m_{8.0}$  & $m_J$  &  $m_H$  &  $m_{K_{s}}$ &  $A_{V}$ &  $M_{J}$  &  $M_{H}$  &  $M_{K_{s}}$ & O-type star \\
                   &  (mag)  &  (mag)  &  (mag)   &  (mag)  &  (mag)  &  (mag)  &  (mag) &  (mag)   & (mag) & (mag) & (mag) &               \\
                   (1)   &  (2)   &  (3)   &  (4)   &  (5)   &  (6)   &  (7)   &  (8) &  (9) &  (10)   &  (11)   &  (12) &  (13)  \\
\hline
G063.0696-00.3941  &  11.34  &  11.39  &  11.29   &  11.23  &  12.63  & 11.89   &  11.60 & 7.06     & -4.04 & -4.02 & -3.86 &  O7V-O6.5V   \\
G063.0819-00.3947  &  11.01  &  10.94  &  10.88   &  10.89  &  12.65  & 11.77   &  11.29 & 9.26     & -4.64 & -4.52 & -4.42 &  O5.5V-O5V   \\
G063.0918-00.3962  &  11.75  &  11.62  &  11.43   &  11.58  &  13.60  & 12.59   &  12.08 & 10.09    & -3.91 & -3.85 & -3.72 &  O7.5V       \\
G063.0958-00.4059  &  11.75  &  11.82  &  11.65   &  11.67  &  13.70  & 12.51   &  12.08 & 10.29    & -3.88 & -3.97 & -3.75 &  O7.5V-O7V   \\
G063.0962-00.3644  &  11.59  &  11.54  &  11.51   &  11.42  &  13.04  & 12.07   &  11.70 & 8.76     & -4.11 & -4.14 & -3.95 &  O7V-O6.5V   \\
G063.1020-00.4375  &  11.55  &  11.60  &  11.48   &  11.43  &  13.51  & 12.32   &  11.87 & 10.45    & -4.12 & -4.18 & -3.98 &  O7V-O6V     \\
G063.1050-00.4143  &  12.16  &  12.16  &  11.96   &  12.26  &  14.20  & 12.95   &  12.53 & 10.51    & -3.43 & -3.57 & -3.32 &  O9V-O8.5V   \\
G063.1080-00.4197  &  10.82  &  10.90  &  10.71   &  10.71  &  12.61  & 11.49   &  11.05 & 10.01    & -4.89 & -4.93 & -4.74 &  O4V-O4V     \\
G063.1118-00.4060  &  10.73  &  10.65  &  10.60   &  10.46  &  12.31  & 11.37   &  10.94 & 9.05     & -4.92 & -4.89 & -4.74 &  O4V-O4V     \\
G063.1127-00.4278  &  12.18  &  12.16  &  11.86   &  12.00  &  14.11  & 12.89   &  12.45 & 10.51    & -3.53 & -3.62 & -3.40 &  O9V-O8.5V   \\
G063.1138-00.4385  &  12.23  &  12.13  &  12.08   &  11.99  &  14.02  & 12.95   &  12.54 & 9.58     & -3.35 & -3.40 & -3.21 &  O9.5V-O9V   \\
G063.1143-00.3893  &  11.09  &  11.13  &  11.02   &  11.04  &  12.43  & 11.57   &  11.26 & 7.81     & -4.44 & -4.47 & -4.29 &  O5.5V-O5V   \\
G063.1168-00.4392  &  10.70  &  10.79  &  10.61   &  10.62  &  12.57  & 11.44   &  11.02 & 9.94     & -4.90 & -4.97 & -4.77 &  O4V-03V     \\
G063.1187-00.4133  &  12.00  &  11.97  &  11.73   &  11.93  &  13.91  & 12.71   &  12.25 & 10.55    & -3.74 & -3.81 & -3.60 &  O8V-07.5V   \\
G063.1251-00.4258  &  10.71  &  10.73  &  10.54   &  10.57  &  12.56  & 11.43   &  11.00 & 10.03    & -4.94 & -5.00 & -4.80 &  O4V-03V     \\

\hline
\end{tabular}
\end{table*}

Toward the bubble N131, in this work, we carried out new
observations of the $J = 1 - 0$ transitions of $^{12}$CO, $^{13}$CO,
and C$^{18}$O using the telescope of the Purple Mountain Observatory
(PMO) at Qinghai province in China. we mainly report several
molecular clumps associated well with the IR dust bubble N131. We
also aim to explore its surrounding ISM and search for star
formation spots. We describe the data in Sect. \ref{sect:data}; the
results and discussion are presented in Sect. \ref{sect:results};
Sect. \ref{sect:summary} summarizes the results.

\section{Observation and data processing}
\label{sect:data}

\subsection{The online archive}

The data of the online archive used in this work include GLIMPSE
\citep{benj2003,chur2009}, MIPSGAL \citep{care2009}, the Two Micron
All Sky Survey (2MASS)\footnote{2MASS is a joint project of the
University of Massachusetts and the Infrared Processing and Analysis
Center/California Institute of Technology, funded by the National
Aeronautics and Space Administration and the National Science
Foundation.} \citep{skru2006}, IRAS \citep{neug1984}, and NVSS
\citep{cond1998}. GLIMPSE is an MIR survey of the inner Galaxy
performed with the Spitzer Space Telescope. We used the mosaicked
images from GLIMPSE and the GLIMPSE Point-Source Catalog (GPSC) in
the Spitzer-IRAC (3.6, 4.5, 5.8, and 8.0 $\mu$m). IRAC has an
angular resolution between $1.5''$ and $1.9''$
\citep{fazi2004,wern2004}. MIPSGAL is a survey of the same region as
GLIMPSE, using the MIPS instrument (24 and 70 $\mu$m) on Spitzer.
The MIPSGAL resolution is 6$''$ at 24 $\mu$m. The IRAS Point Source
Catalog consists of 245889 sources found and verified by the IRAS
(InfraRed Astronomy Satellite) at 12, 25, 60, and 100 $\mu$m. The
NRAO VLA Sky Survey (NVSS) is a 1.4 GHz continuum survey covering
the entire sky north of -40$^{\circ}$ declination \citep{cond1998};
the NVSS survey has a noise of about 0.45 mJy beam$^{-1}$.

\begin{figure*}
\centering
\includegraphics[width=0.85\textwidth, angle=0]{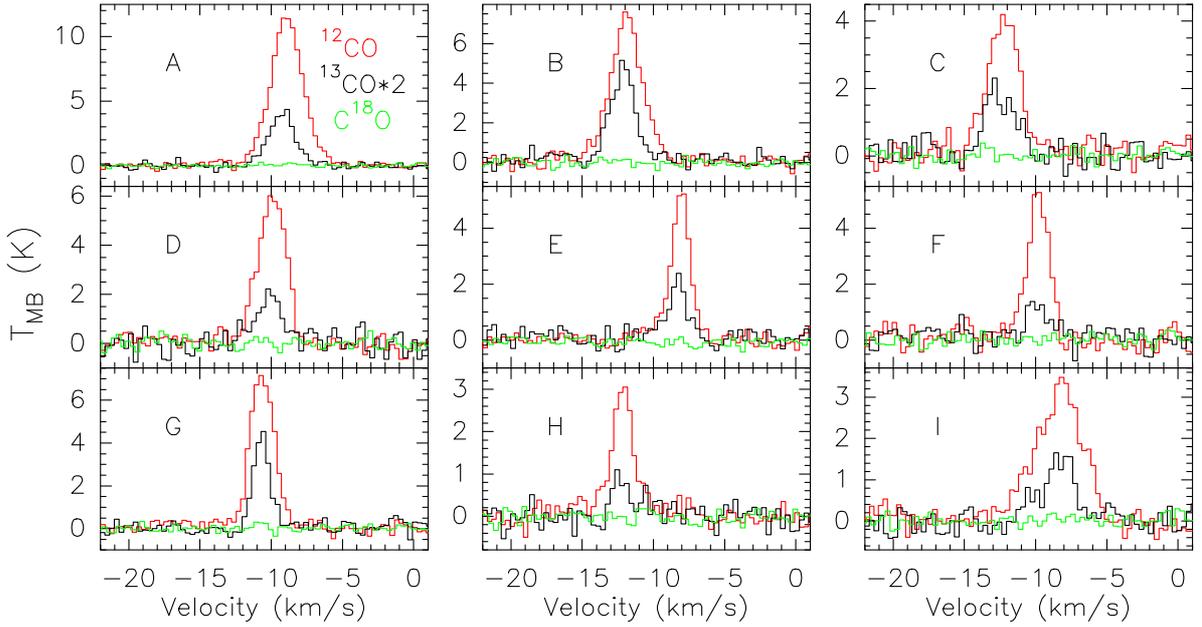}
\caption{$^{12}$CO (red line), $^{13}$CO (black line), and C$^{18}$O
(green line) spectra at the peaks of the molecular clumps from A to
I. The brightness temperature of each $^{13}$CO spectrum is
multiplied by 2.} \label{Fig:core_spectra}
\end{figure*}

\begin{figure*}
\centering
\includegraphics[width=0.75\textwidth, angle=0]{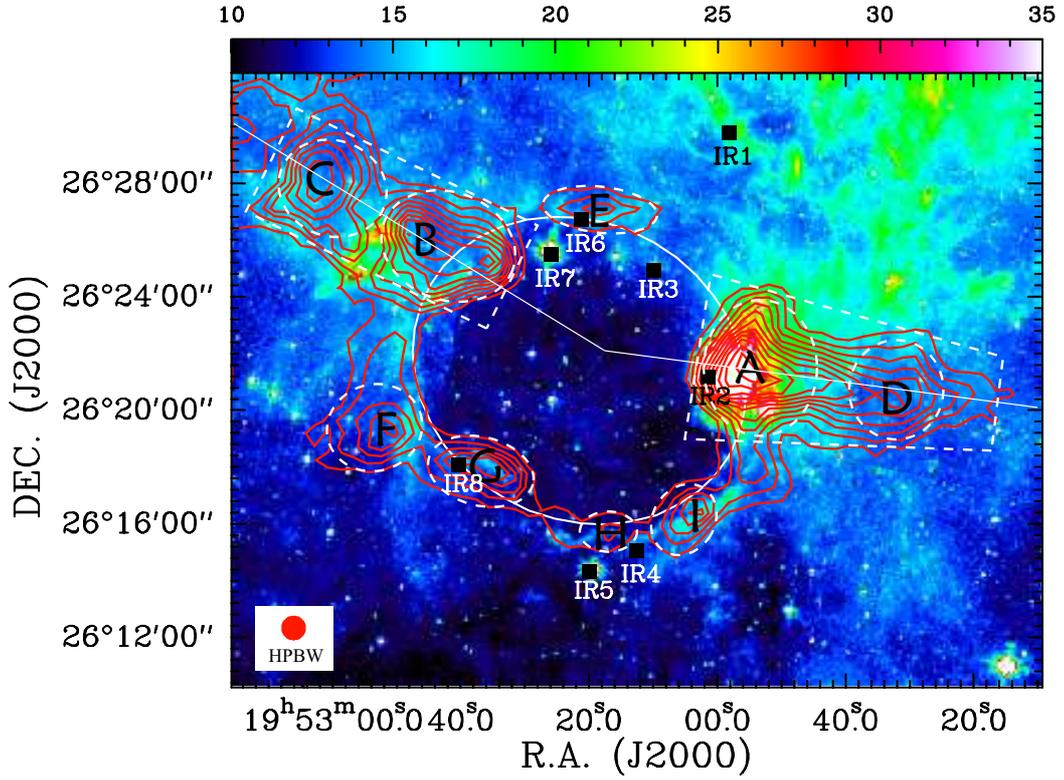}
\caption{Integrated intensity contours of the $^{12}$CO emission
clouds superimposed on the GLIMPSE 8.0 $\mu$m colorscale. The
contour levels range from 5.35 to 32.11 by 2.68 K km s$^{-1}$. The
integration range is from -14.5 to -6.5 km s$^{-1}$. The letters
from A to I indicate the positions of nine molecular clumps, and the
area of each clump is indicated with dashed ellipse and polygon. The
black symbols $``\blacksquare"$ indicate the positions of IRAS point
sources. The white ellipse indicates the position of bubble N131,
and the straight lines indicate the position of position-velocity
diagram in Fig. \ref{Fig:12CO_pv}. The beam size of $^{12}$CO
emission is given in the filled red circle. The unit of the color
bar is in MJy sr$^{-1}$.} \label{Fig:12CO_map}
\end{figure*}

\begin{figure*}
\centering
\includegraphics[width=0.75\textwidth, angle=0]{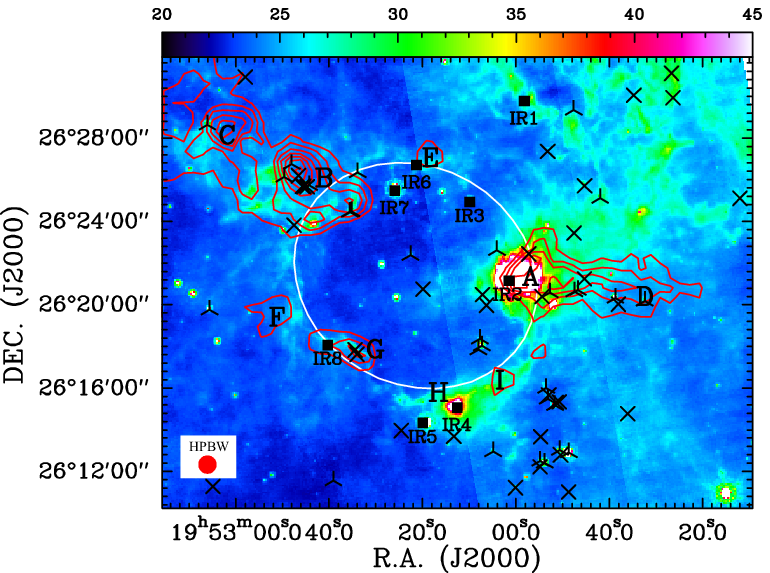}
\caption{Integrated intensity contours of the $^{13}$CO emission
clouds superimposed on the MIPSGAL 24 $\mu$m colorscale. The contour
levels range from 2.17 to 8.68 by 1.09 K km s$^{-1}$. The
integration range is from -14.5 to -6.5 km s$^{-1}$. The letters
from A to I indicate the positions of nine molecular clumps, and the
symbols $``\Yup"$, $``\times"$, and $``\blacksquare"$ indicate the
positions of class I, class II, and IRAS point sources,
respectively. The white ellipse indicates the position of bubble
N131. The beam size of $^{12}$CO emission is given in the filled red
circle. The unit of the color bar is in MJy sr$^{-1}$.}
\label{Fig:13CO_24}
\end{figure*}

\subsection{The CO data of Purple Mountain Observatory}

Our CO observations were made during May 2012 using the 13.7-m
millimeter telescope of Qinghai Station at the Purple Mountain
Observatory at Delingha\footnote{http://www.dlh.pmo.cas.cn/}. We
used the nine-pixel array receiver separated by $\sim$180$''$. The
receiver was operated in the sideband separation of single sideband
mode, which allows for simultaneous observations of three CO J = 1 -
0 isotope variants, with $^{12}$CO in the upper sideband (USB) and
$^{13}$CO and C$^{18}$O in the lower sideband (LSB). The half-power
beam width (HPBW) is 52$''$$\pm$3$''$, and the main beam efficiency
is $\sim$50\% at $\sim$110 GHz. The pointing and tracking accuracies
are better than 5$''$. The typical system temperature during our
runs was around 110 K and varies by about 10\% for each beam. A fast
Fourier transform (FFT) spectrometer was used as the back end with a
total bandwidth of 1 GHz and 16384 channels. The velocity resolution
is about 0.16 km s$^{-1}$ at $\sim$110 GHz.

\begin{figure*}
\centering
\includegraphics[width=0.85\textwidth, angle=0]{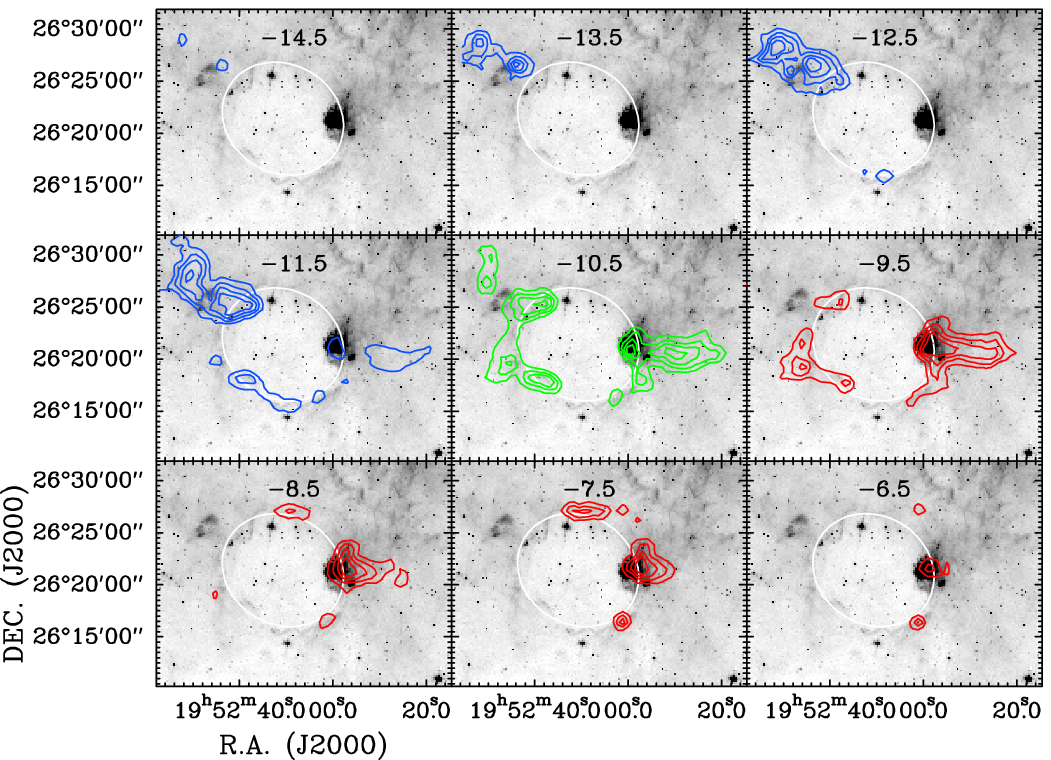}
\caption{Integrated intensity contours of the $^{12}$CO emission
clouds every 1.0 km s$^{-1}$ superimposed on the GLIMPSE 8.0 $\mu$m
grayscale. The lowest contour level for each velocity panel from
-14.5 to -6.5 km s$^{-1}$ is 1.59, 1.37, 1.80, 2.24, 1.96, 2.19,
2.80, 1.22, and 0.95 K km s$^{-1}$, respectively; the level
increment of each panel is equal to the corresponding value of the
lowest contour level. The blue, green, and red contours indicate the
blueshifted, systematic, and redshifted velocities, respectively.
The white ellipse indicates the position of bubble N131.}
\label{Fig:12CO_channel}
\end{figure*}

On-the-fly (OTF) observing mode was applied for mapping
observations. The antenna continuously scanned a region of
20$\arcmin\times20\arcmin$ centered on R.A.(J2000) =
19$^h$52$^m$09$^s$.61, DEC.(J2000) = 26$^{\circ}$22$'$13$''$.4 with
a scan speed of 20$\arcsec$~s$^{-1}$. The OFF position was chosen at
R.A.(J2000) = 19$^h$47$^m$24$^s$.00, DEC.(J2000) =
28$^{\circ}$29$'$24$''$.0, where there is extremely weak CO emission
based on the CO survey of the Milky Way \citep{dame1987,dame2001}.
The rms noise level was 0.2 K in main beam antenna temperature
T$\rm{^{*}_{A}}$ for $^{12}$CO (1-0), and 0.1 K for $^{13}$CO (1-0)
and C$^{18}$O (1-0). The OTF data were then converted to 3-D cube
data with a grid spacing of 30$\arcsec$. The IRAM software package
GILDAS\footnote{http://iram.fr/IRAMFR/GILDAS/} and the software
package MIRIAD\footnote{http://www.cfa.harvard.edu/sma/miriad/} were
used for the data reduction.

\section{Analysis and the results}
\label{sect:results}

\subsection{The dimension and distance of the bubble N131}

We selected the IR dust bubble N131 from the catalog of
\citet{chur2006}. They suggest that the N131 is a complete (closed
ring) IR dust bubble centered on $l^*$ = 63.084, $b^*$ = -0.395,
with an inner short radius r$^*_{in}$ = 5.46$'$, an inner long
radius R$^*_{in}$ = 6.18$'$, and an eccentricity of the ellipse
$e^*_{\rm{N131}}$ = 0.47. By comparing the IR emission with the
integrated intensity of CO emission, the dimensions about the ring
of cloud are respectively r$_{in}$ = 5.20$'$, R$_{in}$ = 6.00$'$,
and $e_{\rm{N131}}$ = 0.50 centered on $l$ = 63.095, $b$ = -0.404,
or R.A.(J2000) = 19$^h$52$^m$21$^s$.5, DEC.(J2000) =
+26$^{\circ}$21$'$24$''$.0. Here, r$_{in}$ and R$_{in}$ are the
semiminor and semimajor axes of the inner ellipse, respectively. By
analyzing H$_{2}$CO absorption (V$_{\rm{H_{2}CO}}$ = 22.6$\pm$0.1 km
s$^{-1}$) against the UC HII region continuum emission
(V$_{\rm{H110\alpha}}$ = -9.3$\pm$2.3 km s$^{-1}$), \citet{wats2003}
resolved the distance ambiguity toward G63.05-0.34, which lies on
the ``far" kinematic distance 8.6$^{+0.9}_{-1.0}$ kpc. In fact,
there is no distance ambiguity toward N131, because the derived
``near" kinematic distance is negative. The G63.05-0.34 is located
at the position of the IRAS 19499+2613, which is correlated with CO
molecular clump A (V$_{\rm{CO}}$ = $\sim$-10.5 km s$^{-1}$) in this
work. \citet{wats2010}, however, adopted the velocity of H$_{2}$CO
($\sim$22.6 km s$^{-1}$) in \citet{wats2003} to obtain a kinematic
distance of $\sim$2.4 kpc toward the bubble N131. Actually, there is
another CO velocity component at $\sim$25.0 km s$^{-1}$, which is
consistent with H$_{2}$CO velocity at $\sim$22.6 km s$^{-1}$, but
not correlated with the ringlike shell of this bubble. This velocity
component of $\sim$22.6 km s$^{-1}$ may belong to the foreground of
the bubble N131. Therefore, we adopt the kinematic distance
D$_{\rm{N131}}$ = 8.6 kpc as the distance of the bubble N131. The
dimensions of the inner short radius and the inner long radius are
D$_{{\rm{r}}_{in}}$ = 13.0 pc and D$_{{\rm{R}}_{in}}$ = 15.0 pc,
respectively.

\subsection{Parameters of the spectra and molecular clumps}

Figure \ref{Fig:core_spectra} shows several spectra for $^{12}$CO,
$^{13}$CO, and C$^{18}$O at the peak locations of nine molecular
clumps from A to I (indicated in Fig. \ref{Fig:12CO_map}) defined on
the basis of $^{13}$CO contours. For the nine molecular clumps, we
detected strong $^{12}$CO and $^{13}$CO emission lines. At any
position toward the N131, however, we did not detect any C$^{18}$O
emission signal of more than 3$\sigma$. We reported spectral
information of the nine molecular clumps in Table
\ref{table:spectra}, and some derived parameters of the molecular
clumps in Table \ref{table:clump}. In Table \ref{table:spectra},
column (1) lists the molecular clump name; columns (2) - (3) list
the equatorial coordinates; columns (4) - (9) list the LSR velocity
V$\rm{_{CO}}$, the full width at half maximum (FWHM)
$\Delta$V$\rm{_{CO}}$, and the peak brightness temperature
T$\rm{_{CO}}$ of each clump using Gaussian fitting for CO emission
line. In Table \ref{table:clump}, columns (1) - (4) show the name,
the dimension, and the integration temperature of each molecular
clump, where the area of each clump is indicated with a dashed
ellipse and polygon in Fig. \ref{Fig:12CO_map}; columns (5) - (8)
show the maximum and mean value of the ionizing temperature and the
H$_{2}$ column density; columns (9) - (10) show the mass and density
of each molecular clump. Here, we assumed each molecular clump is
under the local thermal equilibrium (LTE) assumption, and used the
theory of the radiation transfer and molecular excitation
\citep{winn1979,gard1991}. The excitation temperature T$\rm{_{ex}}$
and column density N$\rm{_{^{13}CO}}$ can be calculated directly,
assuming $^{12}$CO emission to be optically thick and the
beam-filling factor to be unity. The column densities of H$_2$ were
obtained by adopting typical abundance ratios
[H$_2$]/[$^{12}$CO]=10$^4$ and [$^{12}$CO]/[$^{13}$CO]=60 in the
ISM.

\begin{figure*}
\centering
\includegraphics[width=0.8\textwidth, angle=0]{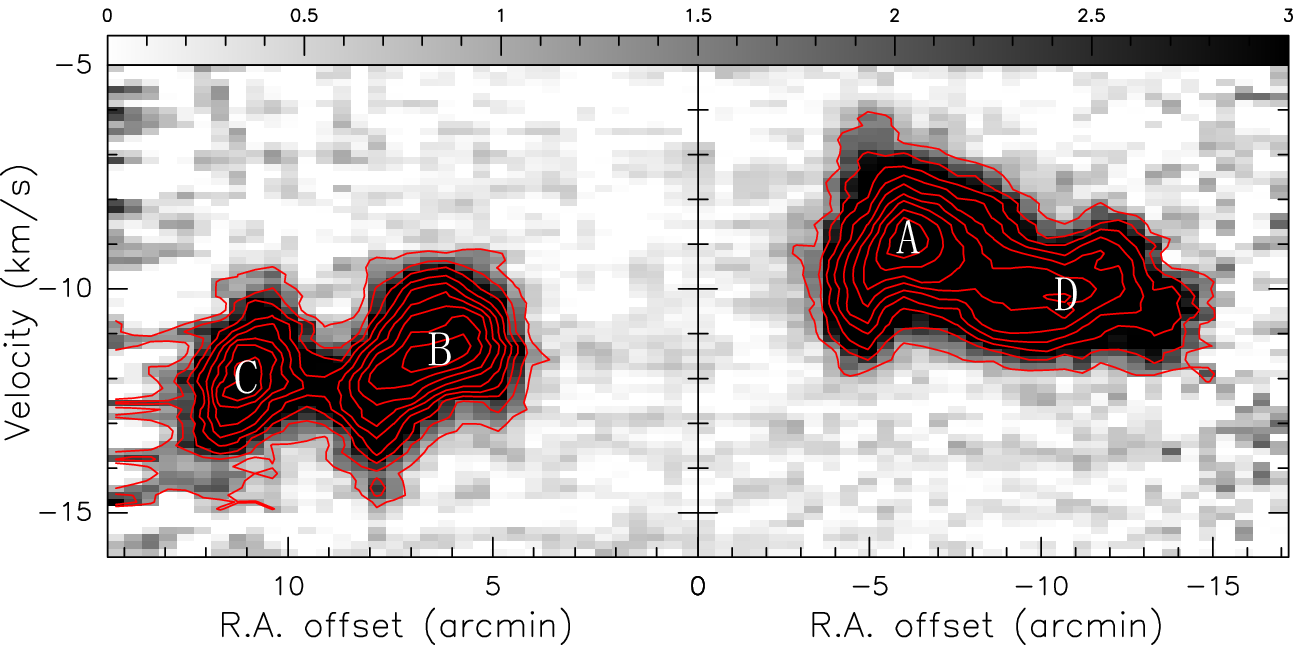}
\caption{The position-velocity diagram of the $^{12}$CO emission
clouds along the white straight lines in Fig. \ref{Fig:12CO_map}.
The position corresponding to offset 0 is at R.A.(J2000) =
19$^h$52$^m$17$^s$.7, DEC.(J2000) = +26$^{\circ}$22$'$05$''$.8. The
position angle of clump BC is positive, and that of clump AD is
negative. The contour levels rang from 0.97 to 8.72 by 0.97 K for
the left panel, while from 1.23 to 11.07 by 1.23 K for the right
panel. The color bar of the grayscale for the $^{12}$CO emission is
above the diagram. The letters A, B, C, and D correspond to the
positions of the same letters as in Fig. \ref{Fig:12CO_map}.}
\label{Fig:12CO_pv}
\end{figure*}

\subsection{The ringlike shell}

\begin{figure*}
\centering
\includegraphics[width=0.75\textwidth, angle=0]{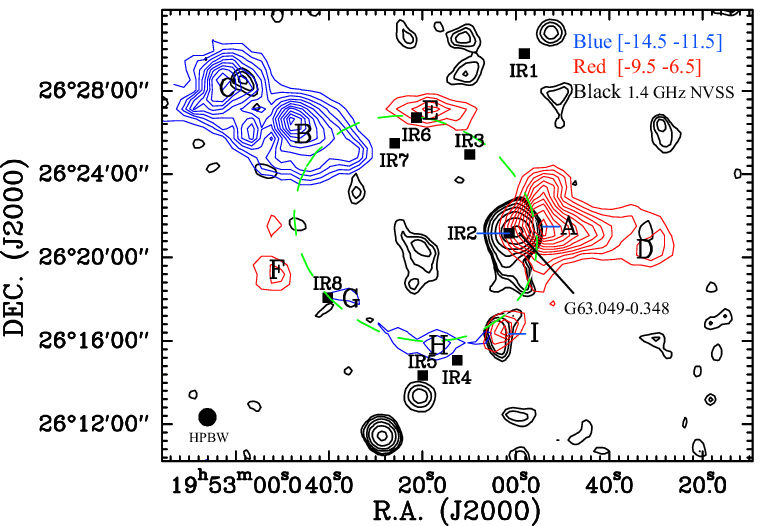}
\caption{Integrated intensity contours of the $^{12}$CO emission of
the blueshifted and redshifted clouds superimposed on the 1.4 GHz
NVSS continuum emission contours. The contour levels range from 3.20
to 19.20 by 1.60 K for the blueshifted cloud, and from 3.86 to 23.17
by 1.93 K for the redshifted cloud. The integration range is from
-14.5 to -11.5 km s$^{-1}$ for blueshifted cloud, and from -9.5 to
-6.5 km s$^{-1}$ for redshifted cloud. The contour levels of the 1.4
GHz NVSS continuum emission are 0.95, 1.26, 1.58, 3.16, 6.32, 12.64,
25.29, and 50.58 mJy beam$^{-1}$. G63.049-0.348 is an HII region.
The letters from A to I indicate the positions of nine molecular
clumps, and the symbols $``\blacksquare"$ indicate the positions of
IRAS point sources. The green ellipse indicates the position of
bubble N131. The beam size of $^{12}$CO emission is given in the
filled black circle.} \label{Fig:12CO_nvss}
\end{figure*}

In Fig. \ref{Fig:12CO_map}, the 8.0 $\mu$m emission colorscale
outlines clearly the distribution of molecular filament and clumpy
structure toward the bubble N131. The 8.0 $\mu$m emission originates
mainly in the polycyclic aromatic hydrocarbons (PAHs). Inside the
N131, the 8.0 $\mu$m emission is much weaker than that at the
ringlike shell. It is likely that the stellar wind from O- and/or
early B-type stars have blown the PAHs onto the ringlike shell. In
Fig. \ref{Fig:13CO_24}, the colorscale is the MIPSGAL 24 $\mu$m
emission. Generally, there should be strong 24 $\mu$m emission and
1.4 GHz continuum emission inside bubble, such as in the dust
bubbles S51 \citep{s51} and N68 \citep{n68}. However, having nearly
the same distribution as 8.0 $\mu$m emission, the 24 $\mu$m emission
is very weak inside the bubble N131.

Comparing $^{12}$CO emission in Fig. \ref{Fig:12CO_map} with
$^{13}$CO emission in Fig. \ref{Fig:13CO_24}, we can find that
several molecular clumps were connected together to form the
ringlike shell of the N131. The inner edge of the ringlike shell has
a much sharper gradient of $^{12}$CO emission than the outer edge.
Based on the distribution of optically thin $^{13}$CO in Fig.
\ref{Fig:13CO_24}, we also found seven clumps A, B, E, F, ..., and I
along the ringlike shell. The parameters of the clumps in Table
\ref{table:clump} show that the dense cores of star formation may be
forming in clumps. Therefore, the ringlike shell is a possible birth
place of star formation that was triggered by the bubble N131.

\subsection{Two elongated molecular clouds}

In Fig. \ref{Fig:12CO_map}, we also found there are two giant
elongated molecular clouds (AD and BC) of CO emission appearing on
opposite sides of the ringlike shell of N131. Morphologically, each
cloud is in alignment, and the central axis of each cloud nearly
goes through the center of the N131. The densest position of each
cloud is located at the ringlike shell of the N131, and the two
clouds AD and BC have extensive structure outwardly. On the north
and south of the clump A located at the ringlike shell, there also
is an expanding tendency in the western direction. This morphology
may be caused by the stellar wind within the bubble, but the
possibility needs to be explored further.

Using the channel map (Fig. \ref{Fig:12CO_channel}) of $^{12}$CO
emission to investigate the velocity component of the bubble N131,
The green contours in the panel of Fig. \ref{Fig:12CO_channel} show
that the systematic velocity of the bubble is about -10.5$\pm$0.5 km
s$^{-1}$, which is consistent with the velocity of the ionized gas
(V$_{\rm{H110\alpha}}$ = -9.3$\pm$2.3 km s$^{-1}$,
\citet{wats2003}). The velocity range of the cloud AD is from -9.5
to -6.5 km s$^{-1}$ indicated with red contours, while the velocity
of the cloud BC is from -14.5 to -11.5 km s$^{-1}$ indicated with
blue contours. We argue that the clouds AD and BC are, respectively,
redshifted and blueshifted relative to the ringlike shell of the
bubble.

In addition, along the central axis of the clouds AD and BC in Fig.
\ref{Fig:12CO_map}, we made a position-velocity diagram in Fig.
\ref{Fig:12CO_pv}. Both the velocity gradient and velocity
dispersion gradient are obvious from clumps A (B) to D (C). The
velocity of the clump A on the ringlike shell is higher than that of
the clump D, which is consistent with the redshifted velocity
distribution of the bubble N68 \citep{n68}. For the cloud BC, the
position-velocity diagram shows that clumps B and C may be two
independent components. However, it is also possible that the
molecular clumps B and C may be interacting. Figure
\ref{Fig:12CO_nvss} shows integrated intensity contours of the
$^{12}$CO emission of the blueshifted (from -14.5 to -11.5 km
s$^{-1}$) and redshifted (from -9.5 to -6.5 km s$^{-1}$) clouds.
There is almost no overlap region between the blue- and redshifted
clouds.

\subsection{Lyman continuum flux}\label{sect:nvss}

\begin{figure*}
\centering
\includegraphics[width=0.49\textwidth, angle=0]{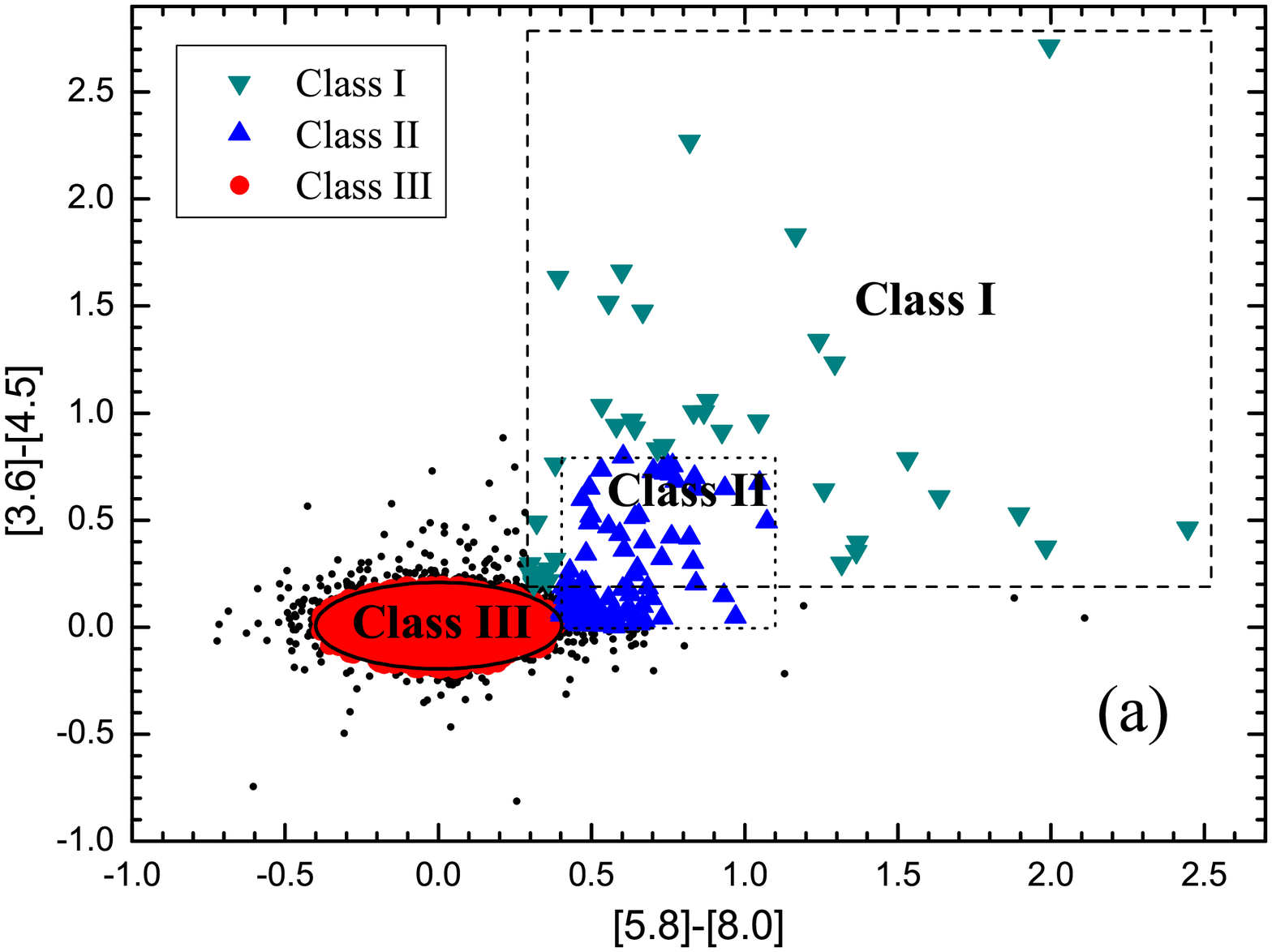}
\includegraphics[width=0.49\textwidth, angle=0]{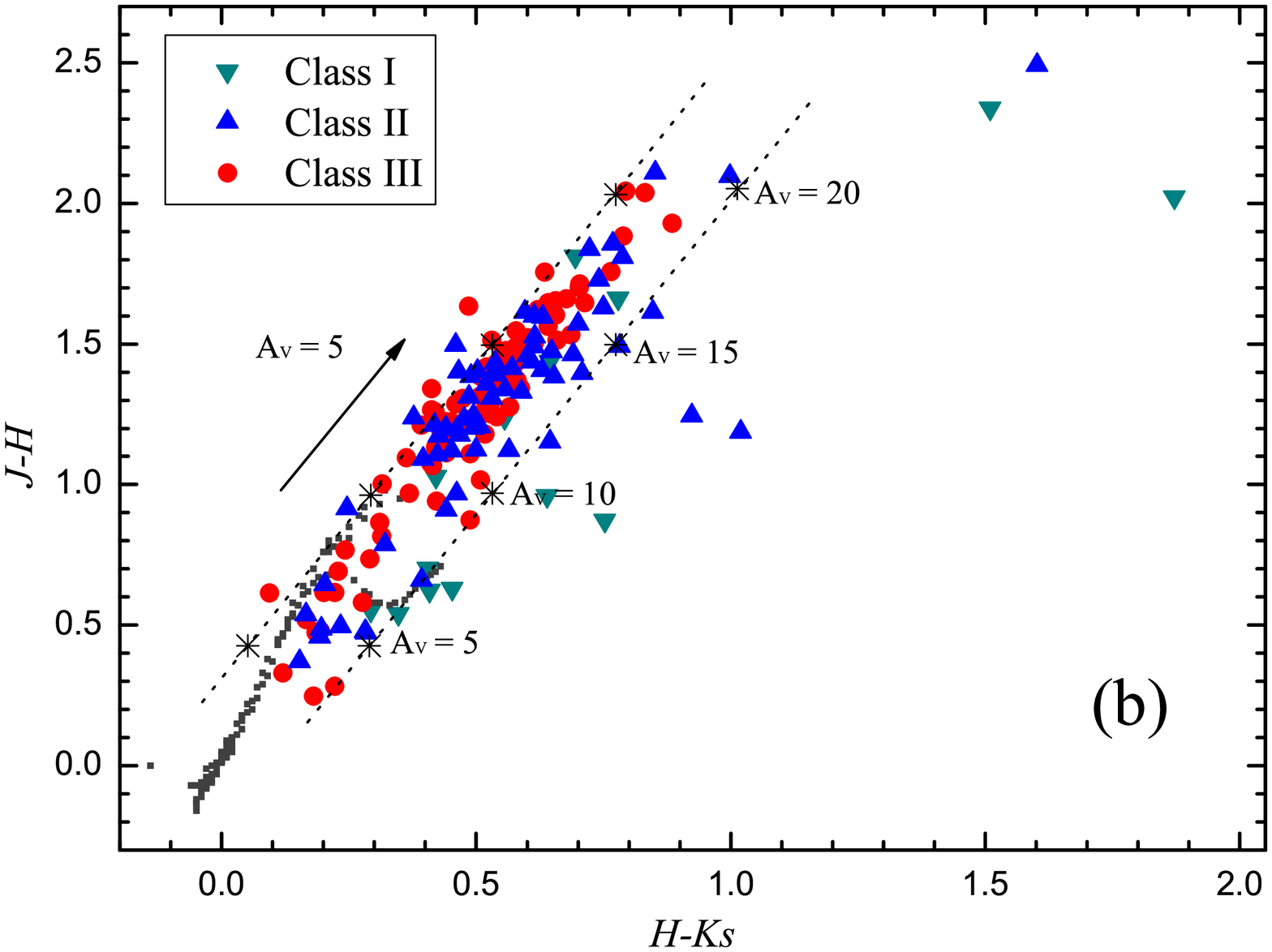}
\caption{(a) GLIMPSE CC diagram [5.8]-[8.0] versus [3.6]-[4.5] for
sources within a circle of about 10$'$ in radius centered on N131.
The classification of class I, II, and III indicates different
stellar evolutionary stages as defined by \citet{alle2004}. (b)
2MASS CC diagram $(H-K_{s})$ versus $(J-H)$. The sources for classes
I and II are these, detected simultaneously by $J$, $H$, $K_{s}$,
3.6, 4.5, 5.8, and 8.0 $\mu$ bands; the sources for class III are
located within a circle of 2.60$'$ (0.5$\times$r$_{in}$) in a radius
centered on N131. The gray squares represent the location of the
main sequence and the giant stars \citep{bess1988}. The parallel
dotted lines are reddening vectors. The adopted interstellar
reddening law is $A_J/A_V$ = 0.282, $A_H/A_V$ = 0.175, and
$A_{K_{s}}/A_V$ = 0.112 \citep{riek1985}, and the intrinsic colors
$(H-K_{s})_{0}$ and $(J-H)_{0}$ are obtained from \citet{mart2006}.}
\label{Fig:color_color}
\end{figure*}

Figure \ref{Fig:12CO_nvss} shows the 1.4 GHz NVSS continuum emission
contours. The region of the continuum flux value above 2$\sigma$
noise (or $>$ 0.9 mJy beam$^{-1}$) was only integrated to consider
as the reliable HII region candidates. From Fig.
\ref{Fig:12CO_nvss}, it can be seen that there is no radio continuum
emission within N131, except for a clump projected on the plane of
the sky towards the center of the bubble. This fact, along with that
the 24 $\mu$m emission being very weak inside the bubble, would
indicate that hot dust and ionized gas have been evacuated by
stellar wind \citep{wats2009}. We also found that IR2 is located at
the peak of HII region G63.049-0.348 \citep{greg1996}, suggesting
that the IR2 (IRAS 19499+2613) may be the ionizing star of the HII
region.

Assuming that the only radio feature in the center of N131 is
associated with the bubble, we derive the radio flux density of the
HII region in about 0.016 Jy. The flux was estimated by integrating
each pixel for the signals of more than 2$\sigma$ noise from the
NVSS fit image. This value should be taken with caution because the
NVSS survey has not added the flux contribution from large-scale
structures. The number of stellar Lyman photon, absorbed by the gas
in the HII region, follows the relation Eq. (\ref{eq:nlcy}) in
\citet{mezg1974}
\begin{equation}
\begin{rm}
    \label{eq:nlcy}
    \left[\frac{N_{Lyc}}{s^{-1}}\right] = 4.761 \times 10^{48}\cdot a\left(\nu, T_e\right)^{-1} \cdot \left[\frac{\nu}{GHz}\right]^{0.1}\cdot
    \left[\frac{T_{e}}{K}\right]^{-0.45}\cdot \left[\frac{S_{\nu}}{Jy}\right]\cdot \left[\frac{D}{kpc}\right]^2,
\end{rm}
\end{equation}
where $\rm{a(\nu, T_e)}$ is a slowly varying function tabulated by
\citet{mezg1967}, for effective temperature of ionizing star
$\rm{T_{e}} \sim$ 33000 K and at radio wavelengths, $\rm{a(\nu,
T_e)} \sim 1$. Finally, we obtained Lyman continuum ionizing photons
flux log$\rm{N_{Lyc}}$ $\sim$ 46.73 from nebula. Assuming the
ionizing stars belong to the O9.5 star with log$\rm{N_{Lyc}}$ $\sim$
47.84 \citep{pana1973}, and the estimated $\rm{N_{Lyc}}$ is a factor
of 2 lower than the expected ionizing photon flux from these stars
\citep{beau2010}, it is suggested that there should be about 0.16
ionizing star of O9.5 star to ionize the ISM within the bubble N131.

We also used the Effelsberg radio continuum data from the Galactic
plane survey at 2695 MHz (only source component) to derive the Lyman
continuum flux\citep{furs1990}. The beam size of this survey is
about 4.3$'$, rms is 20 mK in brightness temperature, and $T_{B}/S =
2.51 \pm 0.05$ [K/Jy]. Within a circle of 2.60$'$
(0.5$\times$r$_{in}$) in radius centered on N131, the average
brightness temperature is about 0.3 K, so the intensity is about
0.120 Jy. Using Eq. (\ref{eq:nlcy}), we obtained the Lyman continuum
flux log$\rm{N_{Lyc}}$ $\sim$ 47.63. After considering the
underestimated factor 2, we found that there is only about 1.24
ionizing star of O9.5 star to ionize the ISM. Therefore, the
Effelsberg radio continuum intensity is also weak like the 1.4 GHz
NVSS continuum. Maybe it is not enough to ionize the bubble N131 for
the weak continuum intensity.

However, we found 15 reliable ionizing star candidates within the
bubble (see Sect. \ref{sect:ysos}). It is likely that hot dust and
ionized gas have been evacuated by a stellar wind from the clustered
ionizing stars, and this will lead to underestimating the Lyman
continuum ionizing photons flux from the ionizing stars.

\subsection{IRAS point sources}\label{sect:iras}

Within a circle about 10$'$ in radius centered on N131, we found
eight IRAS point sources, indicated in Figs. \ref{Fig:12CO_map},
\ref{Fig:13CO_24} and \ref{Fig:12CO_nvss} with the names IR1, IR2,
..., and IR8. These IRAS sources, except IR1, are distributed around
the several molecular clumps in the line of sight. Especially, IR4,
IR6, and IR8 are correlated well with the molecular clumps A, H, E,
and G, respectively. We obtained some parameters of the IRAS point
sources in Table \ref{table:IRAS}. Columns (1) - (2) list the source
name; columns (3) - (4) list the equatorial coordinates; columns (5)
- (8) list the flux of the 12, 25, 60, and 100 $\mu$m, respectively;
columns (9) - (10) list the derived the infrared luminosity
\citep{caso1986} and dust temperature \citep{henn1990}, which are
expressed as
\begin{equation}
    \label{eq:lir}
L_{IR} = (20.653 \times F_{12} + 7.538 \times F_{25} + 4.578 \times
F_{60} + 1.762 \times F_{100}) \times D^{2} \times 0.30,
\end{equation}
\begin{equation}
    \label{eq:lir}
T_{d} = \frac{96}{(3 + \beta) \rm{ln}(100/60) -
\rm{ln}(F_{60}/F_{100})}.
\end{equation}
In the equations above, $D$ is the distance from the Sun in kpc, and
the emissivity index of dust particles $\beta$ is assumed to be two.
Based on the parameters about the infrared luminosity, dust
temperature in Table \ref{table:IRAS}, the eight IRAS point sources
are probable candidates for young massive stars.

In studying of the occurrence of maser emission from star-forming
regions in the very early stages of evolution, \citet{palu1994}
found the 22 GHz H$_{2}$O maser emission has a flux of less than 3.1
Jy, and \citet{vand1995} did not find any 6.7 GHz methanol maser
toward the IR2 (IRAS 19499+2613). Also, in a study surveying the
occurrence of the 22 GHz H$_{2}$O maser emission from bright IR
sources in star-forming regions, \citet{pall1991} found the peak
flux of H$_{2}$O maser is less than 2.7 Jy toward the IR4 (IRAS
19501+2607) near clump H.

\subsection{Ionizing stars and YSOs}\label{sect:ysos}

The GLIMPSE color-color (CC) diagram [5.8]-[8.0] versus [3.6]-[4.5]
in Fig. \ref{Fig:color_color}(a) shows the distribution of class I,
II, and III stars, which are located within a circle of about 10$'$
in radius centered on N131. Here we only considered these sources
with detection in four Spitzer-IRAC bands \citep{hora2008}. Class I
sources are protostars with circumstellar envelopes; class II
sources are disk dominated objects; and class III sources refer to
young stars above the main sequence (and contracting towards it),
but without accretion characteristics such as H$\alpha$ emission
\citep{alle2004,petr2010}.

The reliable ionizing stars are mainly from the class III candidates
and centrally distributed in a small region within the bubble, so we
only considered the sources within a circle of 2.60$'$
(0.5$\times$r$_{in}$) in radius centered on N131. Furthermore to get
rid of the background and foreground stars, we used the 2MASS CC
diagram $(H-K_{s})$ versus $(J-H)$ in Fig. \ref{Fig:color_color}(b).
Considering the extinction in the Galactic plane as a function of
distance to the Sun, \citet{amor2005} present an average Galaxy ISM
extinction A$_V$ =0.96 mag kpc$^{-1}$ for one model. The adopted
distance of the N131 is 8.6 kpc from the Sun, so we just considered
the sources between the extinction range 10.6 $\sim$ 6.6 mag.
Finally, we obtained 15 reliable ionizing star candidates within the
bubble N131. In Table \ref{table:ionizing}, we report the 15
ionizing candidates within the bubble: column (1) specifies the
GLIMPSE designation; columns (2) - (8) list the magnitude of four
Spitzer-IRAC bands and three 2MASS $JHK_{s}$ bands, respectively;
column (9) lists the ISM extinction A$_V$; the absolute $JHK_{s}$
magnitude and spectral type were derived in columns (10) - (13).

YSOs are mainly distributed around the ringlike shell of the bubble,
so we selected YSO candidates from class I and class II sources in
Fig. \ref{Fig:color_color}(a). We then excluded the potential
sources belonging to the reddening main sequence and giant stars
using the 2MASS CC diagram in Fig. \ref{Fig:color_color}(b). These
(class I and II) sources in Fig. \ref{Fig:color_color}(b) only
include those with simultaneous detection in the four Spitzer-IRAC
bands and three 2MASS $JHK_{s}$ bands. Finally, we found 29 class I
stars and 34 class II stars as YSO candidates, which are indicated
around the bubble N131 in Fig. \ref{Fig:13CO_24}. We can see that
the clustered YSOs are distributed on the clouds AD and BC in the
line of sight, and several YSOs are located near the molecular
clumps G and H on the ringlike shell. This distribution provides
some evidence for star formation triggered by the bubble. Within the
bubble, there are two YSO candidates, which are possibly the
background or foreground stars. We also found that there is a good
correlation between the YSOs and MIPSGAL 24 $\mu$m distribution.

\section{Summary}
\label{sect:summary}

Based on our CO emission observations of 13.7-m PMO telescope,
together with other archival data including GLIMPSE, MIPSGAL, 2MASS,
IRAS, and NVSS, we have studied the ISM around the IR dust bubble
N131. The main results can be summarized as follows.

1. We found that the ringlike shell of the associated CO clouds is
well correlated with the Spitzer 8.0 and 24 $\mu$m emission, and
there are two giant elongated molecular clouds (AD and BC) of CO
emission appearing on opposite sides of the ringlike shell of N131.
The two elongated clouds may be triggered by the stellar wind from
the clustered ionizing stars within the N131, but this possibility
needs to be explored further.

2. Contrasting the morphologic distributions between the CO and 8.0
$\mu$m emissions, we found that the CO radii agree rather well with
those from the 8 $\mu$m image. D$_{\rm{N131}}$ = 8.6 kpc was adopted
as the distance of the bubble N131 from the Sun.

3. We found there is a cavity within the bubble at 1.4 GHz and 24
$\mu$m, indicating that hot dust and ionized gas have likely been
evacuated by stellar wind.

4. Seven IRAS point sources (IR2, IR3, ..., and IR8) are distributed
along the ringlike shell of the bubble N131. IR2 (IRAS 19499+2613)
is located at the peak of HII region G63.049-0.348. IR2, IR4, IR6,
and IR8 are well correlated with the molecular clumps A, H, E, and
G, respectively.

5. We found 15 ionizing stars and 63 YSO candidates. The clustered
YSOs are distributed along the elongated clouds AD and BC, and
several YSOs are located around the clumps G and I. These locations
may be the birth places of star formation triggered by the bubble
N131.

\begin{acknowledgements}
We wish to thank the anonymous referee and editor Malcolm Walmsley
for comments and suggestions that improved the clarity of the paper.
We are grateful to the staff at the Qinghai Station of PMO for their
assistance during the observations. Thanks go to the Key Laboratory
for Radio Astronomy, CAS, for support the operating telescope. This
work was supported by the Young Researcher Grant of the National
Astronomical Observatories, Chinese Academy of Sciences.
\end{acknowledgements}

\bibliographystyle{aa}
\bibliography{references}

\end{document}